\documentclass[a4paper,11pt]{article}
\usepackage[utf8]{inputenc}

\usepackage{framed}
\usepackage{mdframed}

\usepackage{fullpage}
\usepackage[T1]{fontenc}

\usepackage{epsf}
\usepackage{graphicx}
\usepackage{verbatim}
\usepackage{color}	
\usepackage{bbold}
\usepackage{hyperref}

\usepackage[british]{babel}
\usepackage{verbatim}
\usepackage[T1]{fontenc}
\usepackage{lmodern}
\usepackage{lipsum}
\usepackage{booktabs}
\usepackage{caption}
\usepackage{cite}
\usepackage{soul,color}
\usepackage[toc,page]{appendix}

\usepackage{ifpdf}

\usepackage{amsthm, amssymb}
\usepackage[tbtags]{amsmath}
\usepackage{bm}
\usepackage{mathtools}
\usepackage{amstext}
\usepackage{braket}
\usepackage{multirow}

\usepackage[normalem]{ulem}
\usepackage{xcolor}
\usepackage{cancel}

\newcommand\redsout{\bgroup\markoverwith{\textcolor{red}{\rule[0.5ex]{2pt}{0.4pt}}}\ULon}

\begin{document}
\vspace{5mm}
\vspace{0.5cm}

\def\be{\begin{eqnarray}}
\def\ee{\end{eqnarray}}

\def\ba{\begin{aligned}}
\def\ea{\end{aligned}}

\def\ls{\left[}
\def\rs{\right]}
\def\lc{\left\{}
\def\rc{\right\}}

\def\p{\partial}

\def\S{\Sigma}

\def\s{\sigma}

\def\O{\Omega}

\def\a{\alpha}
\def\b{\beta}
\def\g{\gamma}

\def\ad{{\dot \alpha}}
\def\bd{{\dot \beta}}
\def\gd{{\dot \gamma}}
\newcommand{\ft}[2]{{\textstyle\frac{#1}{#2}}}
\def\ib{{\overline \imath}}
\def\jb{{\overline \jmath}}
\def\Re{\mathop{\rm Re}\nolimits}
\def\Im{\mathop{\rm Im}\nolimits}
\def\trace{\mathop{\rm Tr}\nolimits}
\def\rmi{{ i}}

\def\N{\mathcal{N}}

\newcommand{\SU}{\mathop{\rm SU}}
\newcommand{\SO}{\mathop{\rm SO}}
\newcommand{\U}{\mathop{\rm {}U}}
\newcommand{\USp}{\mathop{\rm {}USp}}
\newcommand{\OSp}{\mathop{\rm {}OSp}}
\newcommand{\Symp}{\mathop{\rm {}Sp}}
\newcommand{\Sl}{\mathop{\rm {}S}\ell }
\newcommand{\Gl}{\mathop{\rm {}G}\ell }
\newcommand{\Spin}{\mathop{\rm {}Spin}}

\def\hc{c.c.}

\numberwithin{equation}{section}

\allowdisplaybreaks

\allowbreak


\begin{titlepage}

\thispagestyle{empty}
\begin{flushright}

\end{flushright}
\vspace{35pt}

\begin{center}
	    { \LARGE{
	    O6-plane backreaction 
	    on scale-separated 
	    Type IIA 
	    AdS$_3$ 
	    vacua 
	    }}

		\vspace{50pt}

		{\Large Maxim~Emelin$^{a,b}$, \ Fotis~Farakos$^{a,b}$, \ George~Tringas$^{c}$}

		\vspace{30pt}

{

$^{a}${ Dipartimento di Fisica e Astronomia ``Galileo Galilei''\\ Universit\`a di Padova, Via Marzolo 8,35131 Padova, Italy}
\vspace{15pt}

$^{b}${   INFN, Sezione di Padova \\
Via Marzolo 8, 35131 Padova, Italy}
\vspace{15pt}

$^{c}${   Physics Division, National Technical University of Athens,\\
15780 Zografou Campus, Athens, Greece }

}

\vspace{40pt}

{ABSTRACT}

\end{center}

\vspace{10pt}

We evaluate the backreaction of O6-planes in scale-separated $\text{AdS}_3$ flux vacua of massive Type IIA. 
Using the appropriate flux scaling we show that the corrections to the various background fields and moduli are controlled 
and subleading when going from smeared to localized sources. 
Similarly, the backreaction corrections to the scalar potential are parametrically small in the scale-separation limit, 
assuming always that the near-O6-plane singularities will find a resolution within string theory, 
even in the presence of a Romans mass. 
Our analysis is based on the equations of motion and therefore applies also to the non-supersymmetric vacua.

\bigskip

\vspace*{\fill}
\noindent
\rule{5.8cm}{0.75pt}\\
{\rm \footnotesize E-mails: maxim.emelin@pd.infn.it, fotios.farakos@pd.infn.it, georgiostringas@mail.ntua.gr}

\end{titlepage}


{\hypersetup{hidelinks}
\tableofcontents
}

\setcounter{footnote}{0}

\baselineskip 5.6 mm


\pagebreak

\section{Introduction}

The low-energy dynamics of string theory compactifications can be described by lower-dimen- sional 
effective theories whose properties are determined by the specifics of the internal geometry, fluxes and other ingredients. 
The range of effective theories that can be obtained in this fashion is vast, 
but it appears that not all otherwise internally consistent lower-dimensional effective theories can appear as low-energy limits of string compactifications. 
The delineation of criteria that determine whether an effective theory can be realized in string theory 
(or be consistent with quantum gravity more generally) has come to be known as the swampland program, 
with theories that fail to satisfy these criteria said to reside in the swampland.

One general expectation that has come out of the swampland program is that compactifications 
to non-supersymmetric anti-de Sitter space should be able to decay \cite{Ooguri:2016pdq}, 
and that supersymmetric compactifications cannot have an arbitrarily small internal mean radius compared to the external AdS radius \cite{Lust:2019zwm}. 
In contrast to these conjectures, 
the effective theories describing specific compactifications to non-susy AdS constructed in the literature 
appear to be both fully stable and also enjoy a separation of scales when the supersymmetry breaking effects 
are switched off \cite{Kachru:2003aw,DGKT,Camara:2005dc,LVS,Narayan:2010em}. 
This discrepancy has motivated the further scrutiny of such constructions. 
For instance, 
the supersymmetric vacua appearing in \cite{Kachru:2003aw} 
have been further analyzed and partially challenged 
in a series of publications \cite{Hamada:2019ack,Carta:2019rhx,Gautason:2019jwq}, 
where the gaugino condensation backreaction is properly taken into account as proposed in \cite{Hamada:2018qef}. 
With regards to the constructions in \cite{DGKT}, which are classical, 
one could suspect that the inconsistent approximation is the use of ``smeared'' orientifold sources. 
In the meantime, various difficulties for achieving scale separation in Type II are discussed in \cite{Tsimpis:2012tu,Petrini:2013ika,Gautason:2015tig}, 
further recent developments can be found in \cite{Font:2019uva,Emelin:2020buq,Buratti:2020kda,Lust:2020npd,Tsimpis:2022orc}, 
and implications on the holographic side are discussed in \cite{Conlon:2021cjk,Collins:2022nux,Apers:2022tfm}.

The consistency of the smearing approximation is often challenged and it is also the central subject of this work. 
On one hand, 
the resulting internal manifolds and localized source configurations are some highly complicated solutions of the full higher-dimensional equations of motion, 
whose explicit construction is prohibitively difficult. 
On the other hand, below the compactification scale, 
one could expect the lower-dimensional effective theory to be somewhat insensitive to the local details of the internal manifold, 
at least to leading order in the compactification scale. 
For this reason, one expects to obtain the same lower-dimensional effective description from a ``smeared'' solution, 
in which the charge density from the ``localized'' sources is distributed in a continuous fashion over the internal manifold. 
These solutions are much easier to construct explicitly with many examples in the literature, 
and in some cases the approximation is controllable \cite{Blaback:2010sj,Saracco:2012wc,Baines:2020dmu}. 
More importantly, 
this logic also suggests that properties of the true ``unsmeared'' solution are encoded in higher order corrections in some appropriate perturbative expansion. 
Explicit procedures for finding such an expansion and computing leading corrections to the internal geometry have 
been only recently proposed \cite{Junghans,Marchesano:2020qvg}, 
and applications of this procedure have already appeared in \cite{Cribiori,Marchesano:2021ycx}.

Due to the intricacies of the four-dimensional constructions that exhibit scale separation, 
a much simpler three-dimensional construction with scale separation and smeared sources was put forward in \cite{AdS3}. 
The main motivation behind such work is to use it as an accurate but simpler testing ground for the study of smeared sources, 
but also of other aspects of the swampland program \cite{dS3,Emelin:2021gzx,Apers:2022zjx}. 
In this work we go one step further and we apply the procedure proposed in \cite{Junghans} to the AdS$_3$ vacua of \cite{AdS3}. 
We evaluate the backreaction of the localized sources and we explicitly verify the parametric control over the corrections in the scale-separated limit. 
Our analysis indicates that, assuming such AdS$_3$ solution with localized sources exists, 
the smeared source approximation captures useful information about it - at least to leading order in the backreaction. 
Such assumption has of course the caveat that one has to assume that the O6-plane singularities we encounter here can be resolved within string theory.

\section{Unsmearing the sources} 
\label{section2}

\subsection{The setup}

We start from the bosonic part of the Type IIA supergravity action in the {\it string} frame
\begin{align}\label{actionTypeII}
S_{\text{IIA}}&=\frac{1}{2\kappa^2_{10}}\int \text{d}^{10}X\sqrt{G}\Big( \tau^{2}({\cal R}_{10}-\frac{1}{2}\vert H_3\vert^2)+4G^{MN}\partial_M\tau\partial_N\tau-\frac{1}{2}\vert F_{p}\vert^2\Big) \, ,
\end{align}
where $2\kappa^2_{10}=(2\pi)^7\alpha^{\prime 4}$,
the redefined dilaton field $\tau=e^{-\phi}$, the ten-dimensional determinant of the metric $G\equiv \text{det}(G_{MN})$ and $\vert F_p\vert^2=\frac{1}{p!}F_{\mu_1...\mu_p}F^{\mu_1...\mu_p}$. 
For the local sources we write down only the DBI part in the effective action which is relevant for our analysis and we ignore the Chern–Simons terms and the fluctuations of world-volume fields of the Dp-branes. The Chern–Simons terms will of course be properly taken into account when we check the Bianchi identities/tadpole conditions. For the contribution of the localized sources to the effective action we thus have 
\begin{align}\label{actionOpDp}
    S_{\text{Op}/\text{Dp}}&= - T_p \int \text{d}^{10}X \sqrt{G}\sum_i\tau\delta(\pi_i)   \, ,
\end{align}
where $\delta(\pi_i)$ is a unit-normalized delta-like distribution denoting the locus of the sources that wrap the cycle $\pi_i$. 
For example, 
in our three-dimensional compactification, 
for a space-filling O6-plane $\pi_i$ refers to four-cycles and $\tilde \pi_i$ to 3-cycles. 
The coefficient $T_p$ is given by 
\be 
T_p=N_{\text{Op}}\mu_{\text{Op}}+N_{\text{Dp}} \mu_{\text{Dp}} \,, 
\ee 
and denotes total tension of all the sources wrapping a given cycle and the individual D-brane and O-plane tensions are given by
\begin{align}
&\mu_{\text{Dp}}=(2\pi)^{-p}(\sqrt{\alpha^{\prime }})^{-(p+1)} \, ,   \\
&\mu_{\text{Op}}=-2^{p-5}\times \mu_{\text{Dp}} \, .
\end{align}
It is important to stress that the reason we have $N_{\text{Op}}$ and $N_{\text{Dp}}$ appearing is 
because the delta-distributions $\delta(\pi_i)$ integrate to unit.

We will be interested in a flux background where the external space is (warped) AdS$_d$ and the internal space is compact. 
To this end we make an ansatz for the ten dimensional metric, always in the string frame, of the form 
\begin{align}\label{metric}
    \text{d}s^2_{10}&=w^2(y)g_{\mu\nu}\text{d}x^{\mu}\text{d}x^{\nu}+g_{mn}\text{d}y^{m}\text{d}y^{n} \,, 
\end{align}
where $g_{\mu\nu}$ is the unwarped $d$-dimensional external metric and $g_{mn}$ is the $(10-d)$ dimensional internal one. 
For convenience in computing the stress tensor, we write the local sources in terms of the ten dimensional metric. 
Note however that, for external spacetime filling sources, the action can be expressed in terms of the source worldvolume metric by using the relation 
\begin{align}
    \delta(\pi_i)\equiv \frac{\sqrt{g_{\pi_i}}}{\sqrt{g_{(10-d)}}}\delta^{(9-p)}(y) \, ,
\end{align}
where $g_{\pi_i}\equiv\text{det}((g_{\pi_i})_{\alpha\beta})$ is the metric determinant of the wrapped cycle. The $\delta^{(9-p)}(y)$ function collectively denotes the localized positions of the sources in the internal space and integrates to one 
\be 
\int_{\tilde{\pi}_i}\, \text{d}^{9-p}y\delta^{(9-p)}(y)=1 \,, 
\ee 
over the dual cycle $\tilde{\pi}_i$.

\subsection{Equations of motion}

To compare the localized solutions to the smeared ones and find the next to leading order corrections we first need to calculate the equations of motions for the metric, dilaton and the fluxes. 
We start with a general discussion including all the possible sources in the equations, eventually restricting to the specific choice of sources and fluxes that we are interested in.

\subsubsection{Equations with localized sources}

We set will now set 
\be 2\pi\sqrt{\alpha^{\prime}}=1 \,, 
\ee 
and write down the equations of motion for the fluxes together 
with their sources which are given by the Bianchi identities. 
For sources that wrap cycles of the internal space we have 
\begin{align}\label{deltarelation}
    \int_{\pi_i}\text{vol}_{\pi_i}=\int\text{vol}_{\pi_i}\wedge \delta_{i,9-p}=\int \text{d}^{10-d}y\sqrt{g_{10-d}} \, \delta(\pi_i)\,, 
\end{align}
where $\text{vol}_{\pi_i}$ is the volume density of the wrapped $\pi_i$ cycles, 
and $\delta_{i,9-p}$ is a {\it unit-normalized} $(9-p)$-form with legs transverse to the sources wrapping the $i$-th cycle and with support on the source locus 
\be
\delta_{i,9-p} = \delta(\pi_i) \,  {\rm d}^{9-p}y_{\perp} \,, 
\ee
where $y_{\perp}$ are the coordinates transverse to the sources and wedge products are implied. For the massive Type IIA supergravity considered in \cite{AdS3} the relevant Bianchi identities, 
including the number of sources wrapping each cycle, 
are 
\begin{align}
    \text{d}F_2&=H_3\wedge F_0 - 2N_{\text{O6}}\sum_i^7\delta_{i,3}+N_{\text{D6}}\sum_i^7\delta_{i,3}\, \label{tadpole1},\\
    \text{d}F_4&=H_3\wedge F_2\,,\label{tadpole3}\\
    \text{d}F_6&=H_3\wedge F_4 - 2^{-3}N_{\text{O2}}\delta_{7}+ N_{\text{D2}}\delta_{7} \,.  \label{tadpole2}
\end{align}
Here $N_{\text{O6/O2}}=0$ if there are no O-planes, 
otherwise it is non-vanishing and depends on the number of fixed points the relevant orientifold involution has in the internal manifold. 
Our specific case will involve $N_{\text{O2}}=2^7$ for the total ``number'' of O2-planes. 
For O6-planes we will have $N_{\text{O6}}=2^3$ for each three-cycle, which are in fact all images of a single O-plane under the G2 orbifold.

In order to proceed further, 
we need the equations of motion for the dilaton and the metric; 
we have performed this analysis for $d$-external dimensions in the Appendix \ref{Ap1a}. 
Note that throughout this work we assume that the dilaton profile does not depend on the external space coordinates but only on the internal ones: $\tau(y)$. 
The equations of motion for the dilaton are found in Eq.(\ref{dilatonD}), and for three external dimensions ($d=3$) they become
\begin{equation}\label{dilaton3}
\begin{aligned}
    0=& -8\nabla^2\tau+2\frac{\tau}{w^2}R_3
    -\frac{24}{w}(\partial_mw)(\partial^m\tau)
    -12\frac{\tau}{w}\nabla_m\nabla^mw
    -12\frac{\tau}{w^2}\nabla_mw\nabla^mw \\
    &+2\tau R_{7} -\tau\vert H_3\vert^2 +2\mu_{6}\sum_i \delta(\pi_i)+2^{-3}\mu_{2} \delta(\pi)\,, 
\end{aligned}
\end{equation}
where we use the notation 
\be 
\mu_6=N_{\text{O6}}-2^{-1}N_{\text{D6}} \ , \quad  \mu_2=N_{\text{O2}}-2^{3}N_{\text{D2}} \,. 
\ee 
To find the variation with respect to the metric we need the stress-energy tensor of the localized sources in the internal space, given by the projector
\begin{align}
    \Pi_{i,mn}=-\frac{2}{\sqrt{g_{\pi_i}}}\frac{\delta \sqrt{g_{\pi_i}}}{\delta g_{mn}}=(g_{\pi_i})^{\alpha\beta}\frac{\partial y^l}{\partial \xi^{\alpha}_i}\frac{\partial y^p}{\partial \xi^{\beta}_i}g_{ml}g_{np} \label{projector1}\, , 
\end{align}
where $\xi_i^\alpha$ are worldvolume coordinates of the branes/planes wrapping the $i$-th cycle.
The Einstein equation in Eq.(\ref{Einstein1D}) becomes
\begin{equation}\label{Einstein13}
\begin{aligned}
    0=& -\frac{\tau^2}{w^2}R_{3}+3\tau^2\Big(w^{-1}\nabla^2w+2w^{-2}\nabla_mw\nabla^mw \Big) +\frac{9}{4}\frac{\tau}{w}\partial_mw\partial^m\tau+\frac{3}{4}\tau\nabla^2\tau + \frac{3}{4}(\partial\tau)^2 \\
    & -\frac{3}{8}\tau^2\vert H_3\vert^2-\frac{3}{2}\sum_{p=0}^6\frac{p-1}{8}\vert F_n\vert^2 +\frac{3}{8}\mu_6\tau\sum_i \delta(\pi_i)+\frac{15}{8}2^{-3}\mu_2\tau \delta(\pi)\,.
\end{aligned}
\end{equation}
The trace-reversed Einstein equations using Eq.(\ref{tracedD}) and Eq.(\ref{RicciTensorInternal}) become
\begin{equation}\label{traced3}
\begin{aligned}
   0=&-\tau^2R_{mn}
   +3\frac{\tau^2}{w}\nabla_m\partial_n w
   +\frac{3}{4}\frac{\tau}{w}g_{mn}(\partial w)(\partial\tau) 
   +\frac{1}{4}g_{mn}\tau\nabla^{2}\tau  \\
   & +\frac{1}{4}g_{mn}(\partial\tau)^2
   +2\tau\nabla_m\partial_n\tau
   -2(\partial_m\tau)(\partial_n\tau)  \\
   &+\frac{1}{2}\tau^2\Big(\vert H_3\vert_{mn}^2 -\frac{1}{4}g_{mn}\vert H_3\vert^2\Big) +\frac{1}{2}\sum_{p=0}^6\Big(\vert F_p\vert_{mn}^2 -\frac{p-1}{8}g_{mn}\vert F_p\vert^2 \Big) \\
   &+\mu_6\sum_i\Big(\Pi_{i,mn}-\frac{7}{8}g_{mn}\Big)\tau \delta(\pi_i)-2^{-4}\mu_2\frac{3}{8}g_{mn}\tau \delta(\pi)\,, 
\end{aligned}
\end{equation}
for $\vert F_p\vert^2_{mn}=\frac{1}{(p-1)!}F_{m\mu_2...\mu_p}F_n^{\mu_2...\mu_p}$. 
Since in the case of our interest O2/D2 sources fill the external space, the projector $\Pi_{mn}$ for them is zero.

\subsubsection{Smearing the sources}

Having found the localized equations of motion we can directly find the ones in the smeared approximation. In this approximation the sources take the form
\begin{align}\label{smearedsources}
    \delta(\pi_i)\rightarrow j_{\pi_i}=\frac{\mathcal{V}_{\pi_i}}{\mathcal{V}_7}=\frac{\int_{\pi_i}\text{d}^4y\sqrt{g_{\pi_i}}}{\int \text{d}^7y\sqrt{g_7}} \, ,~~~~~~~
    \delta_{i,3}\rightarrow j_{i,3}=\frac{\text{vol}_{\tilde{\pi}_i}}{\mathcal{V}_{\tilde{\pi}_i}}=\frac{\text{vol}_{\tilde{\pi}_i}}{\int_{\tilde{\pi}_i}\text{d}^3y\sqrt{g_{\tilde{\pi}_i}}} \,,~~~~~~~
\end{align}
with the three-form volume density given by $\text{vol}_{\tilde{\pi}_i}=\sqrt{g_{\tilde{\pi}_i}}\text{d}y^i\wedge\text{d}y^j \wedge\text{d}y^k=e^i\wedge e^j\wedge e^k$, 
where $i,j,k$ are directions transverse to the O6-plane. 
For clarity we note that in our work $\pi_i$ refer to four-cycles and $\tilde \pi_i$ to the corresponding/dual three-cycles; 
we will specify these once we turn to the G2 example. 
Thus each smeared source that enters the Bianchi is normalized with respect to its own three-cycle volume. 
The $\mathcal{V}_7$ is the internal space volume, which we will explicitly define for our example later. 
We accompany the smeared approximation by the following additional assumptions, which will be justified by the equations of motion: 
\begin{itemize}
    \item The warp factor $w(y)$ of the external space as well as the dilaton $\tau(y)$ are slowly varying with respect to the internal coordinates 
    and can be considered to be constant $w(y)\equiv const.$ and $\tau(y)\equiv const.$ 
    \item The background field strengths 
    satisfy $\text{d}F_n = 0 = \text{d} \star F_n$ and similarly for the $H$-flux, 
    and are thus expanded on the harmonic forms of the 7d internal space, 
    while the latter is chosen to be Ricci-flat, that is $R_{mn}=0$\,. 
\end{itemize}
The equations of motion of the dilaton in Eq.(\ref{dilaton3}) in the smeared approximation simplify to
\begin{equation}\label{dilatonsmeared}
    0= 2\frac{\tau}{w^2}R_3 -\tau\vert H_3\vert^2 +2\mu_6\sum_i j_{\pi_i}\,.
\end{equation}
The Einstein equation in Eq.(\ref{Einstein13}) becomes
\begin{equation}\label{EinsteinSmeared}
\begin{aligned}
    0=& -\frac{\tau^2}{w^2}R_{3} -\frac{3}{8}\tau^2\vert H_3\vert^2-\frac{3}{2}\sum_{p=0}^6\frac{p-1}{8}\vert F_p\vert^2 +\frac{3}{8}\mu_6\tau\sum_i j_{\pi_i}\,, 
\end{aligned}
\end{equation}
and the trace-reversed Einstein equations of Eq.(\ref{traced3}) reduce to 
\begin{equation}\label{tracedsmeared}
\begin{aligned}
   0=& \ \frac{1}{2}\tau^2\Big(\vert H_3\vert_{mn}^2 -\frac{1}{4}g_{mn}\vert H_3\vert^2\Big) 
   +\frac{1}{2}\sum_{p=0}^6\Big(\vert F_p\vert_{mn}^2 -\frac{p-1}{8}g_{mn}\vert F_p\vert^2 \Big) 
   \\ 
   & +\mu_6\sum_i\Big(\Pi_{i,mn}-\frac{7}{8}g_{mn}\Big)\tau j_{\pi_i}\,.
\end{aligned}
\end{equation}
Considering our assumptions for the smeared localized objects 
and the harmonic expansion of the field strength forms, 
the smeared Bianchi identities of Eqs.(\ref{tadpole1})-(\ref{tadpole2}) become 
\begin{align}
    0&=H_3\wedge F_0 - 2N_{\text{O6}}\sum_i^7j_{i,3}+N_{\text{D6}}\sum_i^7j_{i,3}\, ,\\
    0&=H_3\wedge F_2\,,\\
    0&=H_3\wedge F_4 - 2^{-3}N_{\text{O2}}j_{7}+ N_{\text{D2}}j_{7}\, . 
\end{align}
Here $j_7$ is the seven dimensional form of the internal space because the O2-planes fill the full three-dimensional external one. 
At this point, we also impose by fiat $F_6 = 0$. When we specialize to the case of G2 holonomy, this will be justified by the absence of six-cycles.

The $F_4$ background flux actually splits into two parts 
\begin{align}
    F_4=F_{4A}+F_{4B} = \sum_i \left(f^i + \hat f^i\right) \Psi_i \,,
\end{align} 
where, 
postponing further details for later, 
we only note that 
\be
\Psi_i = \text{basis of harmonic four-forms of the internal space} \,. 
\ee 
The $H$-flux is also expanded on the harmonic three-forms of the internal space and takes the form 
\be
H_3 = \sum_i h^i \Phi_i \, , \quad \Phi_i = \text{basis of harmonic three-forms of the internal space} \,. 
\ee
The $F_4$ splitting refers to the way the RR-flux wedges with the $H$-flux, that is
\begin{align}\label{tads}
    H_3\wedge F_{4A}\equiv 0\,,~~~~~~ H_3\wedge F_{4B}=  2^{-3}N_{\text{O2}}j_{7}- N_{\text{D2}}j_{7} \, .
\end{align} 
The term $H_3\wedge F_{4A}$ vanishes by construction, leaving the $f^i$ unconstrained, except for quantization conditions. Meanwhile, the second equation can either be satisfied by balancing the fluxes terms against the smeared source terms, or by demanding that $H_3\wedge F_{4B}$ vanish independently by setting $F_{4B}=0$ (or equivalently $\hat{f}^i=0$). In the latter case, we require a net charge cancellation between the D2-branes and O2-planes, i.e. $N_{\text{O2}} = 8 N_{\text{D2}}$. 

The integral of $\text{d}F_6$ over the internal closed manifold is zero 
and the tadpole relation is satisfied 
for fixed ``orientation'' of the $F_{4A}$ flux while at the same time its magnitude remains unbounded. In the case when the D2/O2 cancellation happens, we have 
\begin{align}
\label{tadpole-smeared-scale-sep}
    \int_7\text{d}F_6=0\,, ~~~~~~ \hat f^i=0 \,, ~~~~~~  \sum_i h^i f^i=0 \,, ~~~~~~ \sum_i f^i f^i= \text{free}  \,, ~~~~~~ 0=16-N_{\text{D2}} \,, 
\end{align}
always for properly quantized flux coefficients $h^i$ and $f^i$. Scale separation can be achieved parametrically in the limit of {\it large} $f^i$, that is 
\be
\sum_i f^i f^i \gg 1 \quad \Rightarrow \quad \text{separation of KK and AdS scales}\,,  
\ee
therefore it is not prohibited by flux quantization. 
The appropriate flux quantization can be found in \cite{AdS3} and we do not repeat it here. 
When there is no net D2/O2 cancellation one has to consider the appropriate amount of D2-branes because 
the last equation in \eqref{tadpole-smeared-scale-sep} is altered to $\sum_i h^i \hat{f}^i = 2^{-3}N_{\text{O2}} - N_{\text{D2}}$. 
For the rest of the article we will have 
\be
F_4 \equiv F_{4A} \ , \quad \text{unless otherwise noted,} 
\ee
so that we do not clutter the formulas.

\subsection{Scaling of the fields}

Now we use the \textit{smeared} equations of motion with net D2/O2 cancellation that 
we found in the previous subsection and require each term in the equations to have the same scaling. 
As expansion parameter of the fluxes we use the parameter $n$, 
and we will see that the smeared equations of motion are invariant under its variation. 
The expansion parameter can have a physical interpretation as the vacuum expectation value of some field or flux, 
and it will later serve as our expansion parameter when we evaluate the backreaction. 
The fact that the smeared solution leaves the $n$ undetermined means that we can make it parametrically 
large so that we can have a good control over the corrections.

To start we assume that the metric of the internal space has the following scaling at smeared level
\begin{align}
    g_{mn}\sim n^a \, .
\end{align}
Then we consider the smeared O6-plane sources in Eq.(\ref{smearedsources}) which enter the Bianchi identity 
and the Einstein equations, 
and we find the following scaling
\begin{align}
    j_{\pi_i}\sim \frac{\sqrt{g_4}}{\sqrt{g_7}}~\sim n^{-\frac{3}{2}a} \, ,~~~~~~~
    j_{i3}\sim \frac{\sqrt{g_3}}{\sqrt{g_3}}~\sim n^{0} \, .
\end{align}
We notice that the smeared O6-planes which enter the Einstein equations have the same scaling as the ones in the 4d compactification on a Calabi--Yau \cite{Junghans}. 
This happens because the difference of the dimensions between the wrapped volume and the internal space is the same 
in both cases $j_{\pi_i}\sim \sqrt{g_3}/\sqrt{g_6}\sim \sqrt{g_4}/\sqrt{g_7}$. 
The dual current of the wrapped cycles $j_{i,3}$ is a three-form and therefore it has no scaling because it does not depend on the metric. 
The next step is to consider the dilaton and the Einstein equations of motion as well as the Bianchi identities to find the scaling of the fluxes. 
We will work with the ansatz 
\begin{align}
    F_0 \sim n^c~,~F_4 \sim n^f~,~H_3 \sim n^b ~,~\tau \sim n^t~,~w \sim n^w \, . 
\end{align}
Moreover, 
the square of a form of n-rank has the following scaling 
\begin{align}
    \vert F_p\vert^2=\frac{1}{p!}g^{a_1a'_1}...g^{a_pa'_p}F_{a_1...a_p}F_{a'_1...a'_p}\sim n^{-p\times a}\times n^{2\tilde{k}} \, , 
\end{align}
where the $\tilde{k}$ is the RR or NSNS flux, therefore $\tilde{k}=c,f,b$ for our case. 
Let us first check the Bianchi identities which will define the scaling of the RR and $H$ fluxes. 
From the first Bianchi identity in Eq.(\ref{tadpole1}) we get 
\begin{align}
    b + c = 0 \, ,
\end{align}
since the smeared source in the Bianchi is not scaling. 
The second Bianchi in Eq.(\ref{tadpole3}) does not give us any scaling information, 
and the same goes for the third equation in Eq.(\ref{tadpole2}), 
because the specific combinations of fluxes vanish. 
From equation (\ref{EinsteinSmeared}) we find the following scaling relation 
\begin{align}
 ~2\tau-2w=2\tau-3a+2b=2c=-4a+2f=\tau-\frac{3}{2}a  \, , \label{scaling1}
\end{align}
and from the traced Einstein equations (\ref{tracedsmeared}) we find 
\begin{align}
    2\tau-2a+2b=a+2c=-3a+2f=\tau - \frac{1}{2}a \, . \label{scaling2}
\end{align}
Then the dilaton equations of motion in Eq.(\ref{dilatonsmeared}) give the following scaling relation 
\begin{align}
 t-2w=t-3a+2b=-\frac{3}{2}a \, . \label{scaling3}
\end{align}
Solving (\ref{scaling1})-(\ref{scaling3}) and (\ref{tadpole1}) we get 
\begin{align}
    a\rightarrow -\frac{2}{3}t+\frac{4}{3}w~,~~b\rightarrow -t+w ~,~~ c\rightarrow t-w~,~~f\rightarrow -\frac{1}{3}t+\frac{5}{3}w \, .
\end{align}
We need an extra condition to find the proper scaling and this comes from the Romans mass, $F_0$, 
which has no scaling because it is a quantized constant, thus $c=0$. 
The parametric scaling of the fluxes at smeared/leading order then is 
\begin{align}
   &F_4\sim n \, ,~~ 
    F_0\sim n^0 \,,~~H_3\sim n^0 \,,~~
    \tau \sim n^{\frac{3}{4}} \,,~~
    w\sim n^{\frac{3}{4}} \, ,~~
    g_{mn}\sim n^{\frac{1}{2}}\, , 
\end{align}
which is the same scaling as in \cite{Junghans}. 
Another way to see this scaling would be to impose the dilaton and the warp factor to have the same scaling $n^t=n^w$, 
which would then fix the Romans mass to $c=0$. 
It is gratifying to see that the scaling of the fluxes we found here from analysing the full higher-dimensional equations 
does actually agree with the one found in \cite{AdS3} where the low energy effective theory was instead analyzed.

When the flux $F_{4B}$ is not zero its wedge with $H_3$ has to be cancelled by a non-vanishing O2/D2 charge in the Bianchi identity. 
From the variation of the dilaton, including now the net O2/D2 contribution, we find 
\begin{equation}\label{dilatonsmeared2}
    0= 2\frac{\tau}{w^2}R_3 -\tau\vert H_3\vert^2 +2\mu_6\sum_i j_{\pi_i}+2^{-3}\mu_2 j_{\pi}\,.
\end{equation}
Performing the scaling analysis for the smeared sources we see that $j_{\pi_i}\sim n^{-3a/2}$ and $j_{\pi}\sim n^{-7a/2}$, and requiring the equation to be invariant under the $1/n$ scaling we see that the scaling of the metric as well as the rest of the fields have to be zero. From the Bianchi identity in Eq.(\ref{tads}) and considering the scaling of $H_3\sim n^0$ and $j_7\sim n^0$ we directly see that
\begin{align}
F_{4B}\sim n^0\,. 
\end{align}
We will later discuss the contribution of the O2/D2 and $F_{4B}$ in the potential and see how they affect the smeared potential.

\subsection{Next to leading order equations of motion}

In this subsection we expand the RR, NSNS fields and the warp factor in terms of a scaling parameter $n$, which can be interpreted as tracking the leading order scaling of the $F_{4A}$ flux responsible for the scale separation.
The fields in the smeared approximation are the leading order terms of a $1/n^p$ expansion. We then perform the $1/n^p$ expansion to find the first order equations of motion. The power $p$ for each field, i.e. the scaling rate of the next to leading order terms, is not uniquely dictated by the system of equations we have at our disposal. However, with a proper ansatz we can calculate all the next to leading order RR flux corrections. 
Our ansatz is  
\begin{align}
F_6&=F_6^{(0)}n+F_6^{(1)}n^0 + \mathcal{O}(n^{-1}) \, ,\label{expansion0} \\
F_4&=F_4^{(0)}n+F_4^{(1)}n^{0}+\mathcal{O}(n^{-1}) \, , \label{expansion}\\
F_2&=F_2^{(0)}n^{1/2}+F_2^{(1)}n^{0}+\mathcal{O}(n^{-1/2}) \, , \\
H_3&=H^{(0)}_3n^0 + H^{(1)}_3n^{-1} + \mathcal{O}(n^{-2}) \, , \\
\tau &=\tau^{(0)}n^{3/4}+\tau^{(1)}n^{-1/4}+\mathcal{O}(n^{-5/4}) \, , \label{dilatonexpansion} \\
w &=w^{(0)}n^{3/4}+w^{(1)}n^{-1/4}+\mathcal{O}(n^{-5/4}) \, , \label{warpingexpansion}\\
g_{mn} &=g_{mn}^{(0)}n^{1/2}+g_{mn}^{(1)}n^{-1/2}+\mathcal{O}(n^{-3/2}) \, . \label{metricexpansion}
\end{align}
Starting with the Bianchi identities, we expand the fluxes in Eq.\eqref{tadpole1} at first order and we get 
\begin{align}\label{expp1}
    \text{d}\Big(F^{(0)}_2n^{1/2}+F_2^{(1)}+...\Big)=\Big(H^{(0)}_3+H^{(1)}_3n^{-1}+...\Big)\wedge F^{(0)}_0 - 2\mu_6\sum_i \delta_{i,3}\, ,
\end{align}
where at leading order we recover the smeared expression along with the first order correction of the Bianchi identity
\begin{align}
    &\text{d}F_{2}^{(0)}=0 \, , \\ &\text{d}F_{2}^{(1)}=H_3^{(0)}\wedge F_0^{(0)} - 2\mu_6\sum_i \delta_{i,3} \, . \label{Biancha}
\end{align}
For the Bianchi identity in Eq.(\ref{tadpole3}) we get
\begin{align}\label{expp2}
    \text{d}\Big(F^{(0)}_4n^1+F_4^{(1)}n^{0}...\Big)=\Big(H^{(0)}_3n^0+H^{(1)}_3n^{-1}+...\Big)\wedge \Big(F^{(0)}_2n^{1/2}+F^{(1)}_2n^0+...\Big) \, ,
\end{align}
from which we deduce 
\begin{align}
    &\text{d}F_4^{(0)}=0 \, , \\
    &\text{d}F_4^{(1)}= H_3^{(0)}\wedge F_2^{(1)}\neq 0 \, . \label{dF4}
\end{align}
We notice that the first order correction of this Bianchi contains the RR two-form correction $F_2^{(1)}$ whose exact form is calculated in the next section using the Einstein equations. 
For the Bianchi identity in Eq.(\ref{tadpole2}) we have
\begin{align}
\label{BI-EXP-EXP}
    \text{d}\Big(F_6^{(0)}n+F_6^{(1)}n^0...\Big)=\Big(H^{(0)}_3n^0+H^{(1)}_3n^{-1}+...\Big)\wedge \Big(F^{(0)}_4n+F^{(1)}_4n^{0}+...\Big)-2^{-3}\mu_2\delta_7 \, . 
\end{align}
Considering the case of our interest, 
where both $H_{3}^{0}\wedge F_{4A}^{(0)}$ and $F_6^{(0)}$ vanish in the smeared approximation, we have  
\begin{align}
    &\text{d} F_6^{(0)}=H^{(0)}_3\wedge F^{(0)}_{4A}=0 \, , \\ 
    &\text{d} F_6^{(1)}=H^{(0)}_3\wedge F^{(1)}_{4A}\neq 0 \,. \label{dF6}
\end{align}
At leading order the orientation of the fluxes leads to the desired cancellation, 
while at subleading order we can always set 
\be
\int_7 \text{d}F_6^{(1)}=\int_7 H^{(0)}_3\wedge F^{(1)}_{4A} = 0 \, , 
\ee
by adjusting the harmonic parts of $F^{(1)}_{4A}$ such that no new sources are required for the tadpole cancellation.

We now turn to the first order expression of Einstein and dilaton equations of motion Eq.(\ref{dilaton3})-(\ref{traced3}). The dilaton equation is
\begin{equation}\label{dilaton3FIRST}
\begin{aligned}
    0=& -8\nabla^2\tau^{(1)}+2\frac{\tau^{(0)}}{(w^{(0)})^2}R_3
    -12\frac{\tau^{(0)}}{w^{(0)}}\nabla_m\nabla^mw^{(1)} + 2\tau^{(0)} R_{mn}^{(1)}g^{(0)mn} -\tau^{(0)}\vert H_3^{(0)}\vert^2 \\
    & +2\mu_6\sum_i \delta(\pi_i)\, ,
\end{aligned}
\end{equation}
and the next to leading order expansion of the Einstein equation in Eq.(\ref{Einstein13}) is
\begin{equation}\label{Einstein13FIRST}
\begin{aligned}
    0=& -\frac{(\tau^{(0)})^2}{(w^{(0)})^2}R_{3}+3\frac{(\tau^{(0)})^2}{w^{(0)}}\nabla^2w^{(1)} +\frac{3}{4}\tau^{(0)}\nabla^2\tau^{(1)}  \\
    & -\frac{3}{8}(\tau^{(0)})^2\vert H_3^{(0)}\vert^2-\frac{3}{2}\sum_{p=0}^6\frac{p-1}{8}\vert F_p^{(0)}\vert^2 +\frac{3}{8}\mu_6\tau^{(0)}\sum_i \delta(\pi_i)~.
\end{aligned}
\end{equation}
Next, the first order correction to the trace reversed Einstein equation in Eq.(\ref{traced3}) becomes
\begin{equation}\label{traced3FIRST}
\begin{aligned}
   0=&-(\tau^{(0)})^2R_{mn}^{(1)}
   +3\frac{(\tau^{(0)})^2}{w^{(0)}}\nabla_m\partial_n w^{(1)}
   +\frac{1}{4}g^{(0)}_{mn}\tau^{(0)}\nabla^{2}\tau^{(1)}  +2\tau^{(0)}\nabla_m\partial_n\tau^{(1)}  \\
   &+\frac{1}{2}(\tau^{(0)})^2\Big(\vert H_3^{(0)}\vert_{mn}^2 -\frac{1}{4}g^{(0)}_{mn}\vert H_3^{(0)}\vert^2\Big) +\frac{1}{2}\sum_{p=0}^6\Big(\vert F_p^{(0)}\vert_{mn}^2 -\frac{p-1}{8}g^{(0)}_{mn}\vert F_p^{(0)}\vert^2 \Big) \\
   &+\mu_6\sum_i\Big(\Pi_{i,mn}^{(0)}-\frac{7}{8}g_{mn}^{(0)}\Big)\tau^{(0)} \delta(\pi_i)~.
\end{aligned}
\end{equation}
We combine the smeared and the first order equations of motion to find the following relations 
for the RR, the dilaton and the warping 
\begin{align}
    \text{d}F^{(1)}_2&=2\mu_6\sum_i(j_{i,3}-\delta_{i,3})\,,\label{bianchi-cor1}  
\\
    \nabla^2\tau^{(1)}&=-\frac{3}{2} \mu_6\sum_i(j_{\pi_i}-\delta(\pi_i))\,,\label{bianchi-cor2} 
\\
    \nabla^2w^{(1)}&=\frac{1}{2}\frac{w^{(0)}}{\tau^{(0)}}\mu_6\sum_i(j_{\pi_i}-\delta(\pi_i))\,.\label{bianchi-cor3} 
\end{align} 
For the backreaction on the internal metric we have 
\begin{align}
\tau^{(0)} R_{mn}^{(1)} 
-3\frac{\tau^{(0)}}{w^{(0)}}\nabla_m\partial_nw^{(1)} 
-2\nabla_m\partial_n\tau^{(1)} 
= \mu_6\sum_i\Big( \frac{1}{2} g_{mn}^{(0)} - \Pi_{i,mn}^{(0)}\Big)(j_{\pi_i}-\delta(\pi_i))~.  \label{Ricci2}    
\end{align} 
These equations determine the backreaction of the localized sources on the solution from the smeared approximation 
and can be used in different setups. 
To proceed further we need to work on a specific example therefore we focus on a G2 orientifold.

\section{The G2 orbifold example}
\label{section3}

\subsection{The internal manifold}

So far we have found the formal expressions for the first order corrections to some of the fields using just the presence of O6-planes 
and the dimensions of the internal space. 
To find the exact form of the corrections at first order we need to specify the internal geometry 
and solve Eqs. \eqref{bianchi-cor1}-\eqref{Ricci2}. 
We consider the toroidal orbifold $T^7/(Z_2\times Z_2\times Z_2)$ with periodically identified coordinates of the seven torus 
\begin{align}\label{periodicity}
    y^m\sim y^m + 1\, ,~~~~ m=1,...,7 \,.
\end{align}
The finite group of isometries $\Gamma$ forming the orbifold group preserves the three-form
\begin{align}\label{3form}
    \Phi=e^{127}-e^{347}-e^{567}+e^{136}-e^{235}+e^{145}+e^{246}\,,
\end{align}
where $e^{127}=e^{1}\wedge e^{2}\wedge e^{7}$, etc., 
and here we can also define basis of harmonic three-forms
\be
\label{3basis}
\Phi_i = \left( dy^{127}, - dy^{347}, - dy^{567}, dy^{136}, - dy^{235}, dy^{145}, dy^{246} \right) \ , \quad i = 1 , \dots, 7 \, . 
\ee
Here we have introduced the seven vielbeins of the torus
\begin{align}
    e^m=r^m\text{d}y^m\,,~~~~~~~\text{for}~~ m=1,...,7\,,
\end{align}
while $r^m$ stand for the radii of the corresponding cycles and $\text{d}y^m$ are the orthonormal basis of the internal seven dimensional manifold. 
For completeness it is useful to define the co-associative invariant under $\Gamma$, 
which is a four-form 
\begin{align}
    \star_7\Phi=\Psi=e^{3456}-e^{1256}-e^{1234}+e^{2457}-e^{1467}+e^{2367}+e^{1357}\,, 
\end{align} 
and the basis of harmonic four-forms 
\be
\label{4basis}
\Psi_i = \left( dy^{3456}, -dy^{1256}, - dy^{1234}, dy^{2457}, - dy^{1467}, dy^{2367}, dy^{1357} \right) \ , \quad i = 1 , \dots, 7 \, ,  
\ee
which together with the basis of harmonic three-forms satisfy an orthogonality condition of the form $\int \Phi_i\wedge \Psi_j=\delta_{ij}$. 
The volume of the internal space is 
\begin{align}
\mathcal{V}_7=\prod^7_{m=1}r_m=\frac{1}{7}\int \Phi\wedge\star_7\Phi \,.    
\end{align}
We denote the action of the O2-plane by $\sigma$ and it is $Z_2$ involution on the internal coordinates
\begin{align}\label{Z2}
\sigma :~ y^m \rightarrow  -y^m \, .
\end{align}
Using Eq.(\ref{periodicity}) and Eq.(\ref{Z2}) one can find the loci to be at $\hat{y}^m=0,1/2$. 
The action of the orbifold group $\Gamma=\{\Theta_\alpha,\Theta_\beta,\Theta_\gamma\}$ acts on the torus coordinates in the following way
\be
\begin{aligned}\label{Z2s}
\Theta_\alpha :~ y^m & \to (-y^1, -y^2, -y^3, -y^4, +y^5, +y^6, +y^7) \, , 
\\
\Theta_\beta :~ y^m & \to (-y^1, -y^2, +y^3, +y^4, -y^5, -y^6, +y^7) \, ,
\\
\Theta_\gamma :~ y^m & \to (-y^1, +y^2, -y^3, +y^4, -y^5, +y^6, -y^7) \, . 
\end{aligned}
\ee
The images of the O2-plane under the orbifold involutions are interpreted as O6-planes. 
The positions of the O6-planes are identified with the fixed points of the combined involutions of the orbifold with the $\sigma$. 
In other words the O6-planes are defined by the following involutions 
\begin{align}
\sigma_\alpha = \Theta_{\alpha}\sigma :~ y^m &\to (y^1, y^2, y^3, y^4, -y^5, -y^6, -y^7)\label{fotis} \, ,\\
\sigma_\beta = \Theta_{\beta} \sigma :~ y^m &\to (y^1, y^2, -y^3, -y^4, y^5, y^6, -y^7) \, ,\\
\sigma_\gamma =\Theta_{\gamma}\sigma :~ y^m &\to (y^1, -y^2, y^3, -y^4, y^5, -y^6, y^7) \, , \\
\label{Opr1}
\sigma_{\alpha\beta}= \Theta_\alpha \Theta_\beta \sigma :~ y^m &\to (-y^1, -y^2, y^3, y^4, y^5, y^6, - y^7) \, ,\\
\label{Opr2}
\sigma_{\beta\gamma}= \Theta_\beta \Theta_\gamma \sigma:~ y^m &\to (-y^1, y^2, y^3, -y^4, -y^5, y^6, y^7) \, ,\\
\label{Opr3}
\sigma_{\gamma\alpha} = \Theta_\gamma \Theta_\alpha \sigma:~ y^m &\to (-y^1, y^2, -y^3, y^4, y^5, -y^6, y^7) \, , \\
\label{Opr4}
\sigma_{\alpha\beta\gamma}= \Theta_\alpha \Theta_\beta \Theta_\gamma \sigma:~ y^m &\to (y^1, -y^2, -y^3, y^4, - y^5, y^6, y^7) \, .
\end{align}
The diagonal metric of the internal space and the metric elements can be written as
\begin{align}
    \text{d}s^2_7=\sum_m^7 (r_m)^2\text{d}y^m\text{d}y^m \, , ~~~~~~~ g_{ij}^{(0)}\equiv (r_m^{(0)})^2n^{1/2}~,~~~~i=j=1,...,7 \, .
\end{align} 
For more details on this orbifold, and a series of different applications, see e.g. \cite{AdS3,Emelin:2021gzx,DallAgata:2005zlf}. Considering the $1/n$ expansion form the metric in Eq.(\ref{metricexpansion}) the radii get corrections
\begin{align}\label{radiiexpansion}
    r_m &=r_m^{(0)}n^{1/4}+r_m^{(1)}n^{-1/4}+\mathcal{O}(n^{-3/4}) \, . 
\end{align}

\subsection{Calculation for a single O6-plane}

\subsubsection{Corrections to the RR flux}

In order to proceed and calculate the first order corrections to the fluxes one should solve the equations in Eq.(\ref{bianchi-cor1})-(\ref{Ricci2}) 
for the seven intersected O6-planes in Eq.(\ref{fotis})-(\ref{Opr4}). 
As a first step we solve the equations with the presence of a single O6-plane, 
indicatively we choose the O6$_{\alpha}$-plane with involution given by Eq.(\ref{fotis}) which wraps the four-cycle $\pi_3$. 
We start from the Bianchi in Eq.(\ref{bianchi-cor1}) in order to calculate the RR fluxes and we write it in terms of the internal geometry basis
\begin{align}
    \text{d}F^{(1)}_2&=2\rho_3 \Big(\text{d}y^5\wedge \text{d}y^6\wedge \text{d}y^7\Big) =-2\rho_3 \,\Phi_3 \,,   \label{dF2}
\end{align} 
where the $\rho_{3}$ refers to the appropriate ``backreaction density''. 
For the specific O6-plane Eq.(\ref{fotis}) which wraps the $\pi_3$, 
this backreaction density term is 
\begin{align}
    \rho_{3} 
= \mu_6 \Big{(} j_{\pi_3}-\delta(\pi_3) \Big{)}
= \mu_6 \left\{ 1 
    - \frac1{N_{\rm O6}} \sum_{m\in\{0,1\}}\delta\Big(y^5-\frac{m}{2}\Big)\delta\Big(y^6-\frac{m}{2}\Big)\delta\Big(y^7-\frac{m}{2}\Big) \right\}  \,. 
\end{align} 
To avoid clutter we do not include the subscript ``$3$'' in $\rho_{3}$ in this part because it is always implied. 
As we will verify momentarily, 
an inspection of Eq.(\ref{dF2}) leads us to guess that the $F_2$ is of the form 
\begin{align}\label{F2sol}
    F_2^{(1)}=-2\star_7(\text{d}\beta_3 \wedge \Psi_3)\,. 
\end{align}
Here we have introduced the function $\beta_3\equiv\beta_3(y)$ which as we will see satisfies a Poisson equation and it will be further specified in the next section. 
For the few next steps we suppress the subscript $3$ to avoid clutter. 
Indeed the derivative on \eqref{F2sol} gives 
\begin{align}\label{dF2sol}
    \text{d}F_2^{(1)}&=-2(\nabla^2\beta) \Phi_3 \,,
\end{align}
which can be verified with the following series of steps 
\begin{align}\label{dF2derivation}
    \text{d}(\star_7(\text{d}\beta \wedge \Psi_3)) 
    = \text{d}(\star_7\text{d}(\beta \wedge \Psi_3)) 
    = \star_7 ( (\nabla^2\beta) \Psi_3 ) 
    = \nabla^2\beta (\star_7\Psi_3) 
    = (\nabla^2\beta) \Phi_3 \,.
\end{align} 
This is easily seen by the fact that $\star \text{d} \star \text{d} (\beta \Psi_3) = \nabla^2 (\beta \Psi_3)$. 
Comparing this to Eq.(\ref{dF2}) we get a Poisson equation for $\beta$ that reads 
\begin{align}\label{Poisson}
    \nabla^2\beta=\rho \, .
\end{align}
Similar to \cite{Junghans} the transverse space at each point on the O6-plane is a three-torus. 
Note that because $F_2^{(1)}$ is not closed we do not need to expand it on harmonic cycles. 
From \eqref{F2sol} we see however that $\sigma:  F_2^{(1)} \to - F_2^{(1)}$ so it is odd, 
as it should be, 
and that $\sigma_{\alpha}: F_2^{(1)} \to - F_2^{(1)}$ so it is again odd as it should be under the O6 involutions, 
and finally that $\Theta_{\alpha}: F_2^{(1)} \to F_2^{(1)}$ therefore it is invariant under the orbifold (as it should be). 
The parities under the other orbifold/orientifold involutions can also be checked to be consistent.

Since we have found the explicit form of $F_2^{(1)}$ we are able to calculate the first order corrections to the rest of the RR forms. Using the Bianchi identity in Eq.(\ref{dF4}) the later becomes
\begin{align}
    \text{d}F_4^{(1)}&= H_3^{(0)}\wedge F_2^{(1)} =\text{d}\Big(-2\sum_ih^i\Psi_i\wedge\beta(y)\Big)\,.
\end{align}
This can be seen from the following steps 
\begin{align}\nonumber 
    H_3^{(0)}\wedge F_2^{(1)} &=-\sum_ih^i\Phi_i\wedge \star_7\text{d}\Big(\beta(y)\wedge \sum_j\Psi_j\Big)=\sum_ih^i\Phi_i\wedge \star_7\text{d}\Big(\Psi_i \wedge \beta(y)\Big) \\
\label{dF4derivation} 
    &=-2\sum_ih^i\Psi_i\wedge\text{d}\beta(y) =\text{d}\Big(-2\sum_ih^i\Psi_i\wedge\beta(y)\Big) \,. 
\end{align}
Thus the co-closed part of $F_4$ which appears beyond the smeared approximation is
\begin{align}
    F_4^{(1)}=-2 \beta(y) \sum_i h^i\Psi_i \,. 
\end{align}
Once more, 
for distances far from the source the $F_4^{(1)}$ becomes negligible as expected in the smeared limit and this becomes clear when we calculate the explicit form of $\beta(y)$. Adding the harmonic part we have 
\begin{align}
    F_4^{(1)}= G^i \Psi_i -2 \beta(y) \sum_i h^i\Psi_i \, , 
\end{align}
where $G^i$ is the corrected flux and can be chosen to be 
\begin{align}
    G^i = 2 h^i \int_{\Psi_i}\text{d}^4 y \beta(y)  \,. 
\end{align}
Thus from Eq.(\ref{dF6}) we have 
\be
\begin{aligned}
    \text{d} F_6^{(1)}=H^{(0)}_3\wedge F^{(1)}_4&= H^{(0)}_3\wedge \Big(G^i \Psi_i -2 \beta(y) \sum_i h^i\Psi_i \Big) 
    \\
    &= h^i\Phi_i\wedge \Big(G^j \Psi_j -2 \beta(y) \sum_j h^j\Psi_j \Big) ~, 
\end{aligned}
\ee
which gives $\int \text{d} F_6^{(1)} = 0$.

\subsubsection{Corrections to the dilaton, warp factor and the metric}
So far we used the Bianchi identity of $F_2^{(1)}$ and the internal geometry in order to specify the explicit form of all the first order corrections of the RR fluxes. However we have not found yet the exact first order corrections to the dilaton, the warp factor and the internal metric.

In order to solve Eq.(\ref{Ricci2}) and identify the first order corrections to the remaining fluxes we start from the following definition of the Ricci tensor of the internal space
\begin{align}\label{Ricci1}
R_{mn}^{(1)}&=-\frac{1}{2}g^{(0)rs}\nabla_m\nabla_ng^{(1)}_{rs} + \frac{1}{2}g^{(0)rs}\Big(\nabla_s\nabla_mg^{(1)}_{rn}+\nabla_s\nabla_ng^{(1)}_{rm}\Big) - \frac{1}{2}\nabla^2g^{(1)}_{mn}\,,
\end{align}
and the relation for the Ricci tensor $R_{mn}$ from Eq.(\ref{Ricci2}) we have
\begin{align}\label{Ricci3}
R_{mn}^{(1)}&= \frac{3}{w^{(0)}}\nabla_m\partial_nw^{(1)}+\frac{2}{\tau^{(0)}}\nabla_m\partial_n\tau^{(1)} 
+\frac{1}{\tau^{(0)}}\sum_i\Big( \frac{1}{2} g_{mn}^{(0)}-\Pi_{i,mn}^{(0)}\Big)\frac{\sqrt{g_{\pi_i}}}{\sqrt{g_7}}\rho_i\, , 
\end{align}
where $\rho_i$ refers to the appropriate backreaction density for the $i$-th cycle. 
Combining Eq.(\ref{Ricci1}) and Eq.(\ref{Ricci3}) we get the equation 
\begin{align}\label{fotaras}
    &-\frac{1}{2}g^{(0)rs}\nabla_m\nabla_ng^{(1)}_{rs} + \frac{1}{2}g^{(0)rs}\Big(\nabla_s\nabla_mg^{(1)}_{rn}+\nabla_s\nabla_ng^{(1)}_{rm}\Big) - \frac{1}{2}\nabla^2g^{(1)}_{mn} \\ 
    \nonumber 
    &=
    \frac{3}{w^{(0)}}\nabla_m\partial_nw^{(1)}
    +\frac{2}{\tau^{(0)}}\nabla_m\partial_n\tau^{(1)} 
    +\frac{1}{\tau^{(0)}}\sum_i\Big( \frac{1}{2}  g_{mn}^{(0)} -\Pi_{i,mn}^{(0)}\Big)\frac{\sqrt{g_{\pi_i}}}{\sqrt{g_7}}\rho_i 
    \, .
\end{align}

Focusing now on the 3rd cycle (and again suppressing the subscript on $\rho_3$ and $\beta_3$), 
we write the volume of the four-cycles wrapping the internal space and the current of the smeared source 
\begin{align}
    \mathcal{V}_{\pi_3}=r_1^{(0)}r_2^{(0)}r_3^{(0)}r_4^{(0)}\,,~~~~ j_{\pi_3}=\frac{1}{r_5^{(0)}r_6^{(0)}r_7^{(0)}}\,. 
\end{align}
Next we calculate the Ricci tensor for cases depending on parallel, transverse and mixed leg components. 
For the calculation we make the following assumption
\begin{align}
    g^{(0)11}g_{11}^{(1)}=g^{(0)22}g_{22}^{(1)}=g^{(0)33}g_{33}^{(1)}=g^{(0)44}g_{44}^{(1)}\,,~~~~~~~ g^{(0)55}g_{55}^{(1)}=g^{(0)66}g_{66}^{(1)}=g^{(0)77}g_{77}^{(1)} \,.
\end{align}
First, when both the legs of the Ricci tensor are along the wrapped cycle, the stress-energy tensor in Eq.(\ref{projector1}) gets the simple form
\begin{align}
    \Pi_{3,mn}=(g_{\pi_3})_{mn} \,, 
\end{align}
for $m,n$ the directions of the wrapped four-cycle. 
The O6-plane wrapping the $\pi_3$ is parallel to the directions $y^1,y^2,y^3,y^4$ and the fields $w,\tau$ and $g_{mn}$ 
are sourced by $\delta(y^5-\hat y^5)\delta(y^6-\hat y^6)\delta(y^7-\hat y^7)$ which depend only on the transverse $y^5,y^6,y^7$ directions. 
We label the wrapped directions with with indices $i,j$ 
and investigate first the case where the components are parallel and same, 
the relation Eq.(\ref{fotaras}) gives the following solution 
\begin{align}\label{AEKARA}
\nabla^2g^{(1)}_{ii} =  \frac{(r_i^{(0)})^2 }{r_5^{(0)}r_6^{(0)}r_7^{(0)}\tau^{(0)}}\rho
\,.~~~~~~~~~~ i=j=1,2,3,4 \,. 
\end{align} 
Now let us check the Ricci tensor for transverse and same directions, $R_{kl}$ with $k=l=5,6,7$ 
\begin{equation}\label{WOP}
\begin{aligned}
    &-2g^{(0)11}\partial_5\partial_5g^{(1)}_{11}-\frac{1}{2}g^{(0)55}\partial_5\partial_5g^{(1)}_{55} - \frac{1}{2}\nabla^2g^{(1)}_{55} \\
    &=
    \frac{3}{w^{(0)}}\partial_5\partial_5w^{(1)}
    +\frac{2}{\tau^{(0)}}\partial_5\partial_5\tau^{(1)}
    +\frac{1}{2}\frac{r_{5}^{(0)}}{r_{6}^{(0)}r_{7}^{(0)}\tau^{(0)}}\rho \,.
\end{aligned}
\end{equation}
For one parallel and one transverse direction, $R_{jk}$, the equation is trivially satisfied. 
For two different transverse directions, i.e. $R_{kl}$ with $k \neq l$ , we can work-out for example the case $R_{56}$ which gives 
\begin{align}\label{woopss}
    0=-\frac{3}{w^{(0)}}\partial_5\partial_6w^{(1)}
    -2g^{(0)11}\partial_5\partial_6g^{(1)}_{11} 
    -\frac{1}{2}g^{(0)55}\partial_5\partial_6g^{(1)}_{55}
    -\frac{2}{\tau^{(0)}}\partial_5\partial_6\tau^{(1)} \,. 
\end{align}
From Eq.\eqref{WOP}-\eqref{woopss} we have 
\begin{align}\label{AEKARAS}
\nabla^2g^{(1)}_{kk} = -  \frac{(r_i^{(0)})^2 }{r_5^{(0)}r_6^{(0)}r_7^{(0)}\tau^{(0)}}\rho
\,,~~~~~~~~~~ k=l=5,6,7 \,. 
\end{align}
and we write again the solution of the warp factor and the dilaton but expressed in terms of the $\rho$ source 
\begin{align}
    \nabla^2\tau^{(1)}&=- \frac{3}{2} 
\frac{1}{r^{(0)}_5r^{(0)}_6r^{(0)}_7}\rho \,,\label{KKE1} \\
    \nabla^2w^{(1)}&=\frac{1}{2}\frac{w^{(0)}}{\tau^{(0)}}\frac{1}{r^{(0)}_5r^{(0)}_6r^{(0)}_7}\rho \,. \label{KKE2}
\end{align}
With the use of the same function $\beta(y)$ as in Eq.\eqref{Poisson}, 
and Eqs.({\ref{AEKARA}}), (\ref{WOP}), (\ref{KKE1}) and (\ref{KKE2}), 
we get the relations 
\begin{align}\label{relations}
-\frac{g^{(1)}_{kk}}{r_i^{(0)2}}=
\frac{g^{(1)}_{ii}}{r_i^{(0)2}}=
-\frac{2\tau^{(1)}}{3\tau^{(0)}}=
\frac{2w^{(1)}}{w^{(0)}}=
\frac{1}{r_5^{(0)}r_6^{(0)}r_7^{(0)}}\frac{\beta(y^5,y^6,y^7)}{\tau^{(0)}} \,. 
\end{align}

\subsection{Solution of Poisson equation}

To solve the Poisson equation in Eq.(\ref{Poisson}) we mostly follow the steps of \cite{Junghans}. 
We introduce a formal solution in terms of Fourier series and estimate the backreaction, 
without specifying the regularization, 
because it is in any case independent of the choice.

We start from the $\beta_3$ and we suppress the subscript $3$ for now as usual and we also take into account that $\mu_6=N_{\rm O6}=8$. 
This means we have to solve the equation 
\be
\label{P2@}
\nabla^2 \beta = 8 -  \sum_{m,n,p\in\{ 0,1\}}\delta\Big(y^5-\frac{m}{2}\Big)\delta\Big(y^6-\frac{n}{2}\Big)\delta\Big(y^7-\frac{p}{2}\Big) \,. 
\ee
To solve this we expand $\beta$ as 
\be
\beta = \sum_{m,n,k\in\{ 0,1\}} \phi_{mnp} \, , 
\ee
where 
\be
\nabla^2 \phi_{mnk} =  1 - \delta\Big(y^5-\frac{m}{2}\Big)\delta\Big(y^6-\frac{n}{2}\Big)\delta\Big(y^7-\frac{p}{2}\Big)  \,. 
\ee
We first look at one of the fixed points and use the Fourier transform of the delta distribution to get 
\be
1 - \delta(y^5)\delta(y^6)\delta(y^7) 
= 1-\sum_{\Vec{k}\in \mathbb{Z}^3} e^{2\pi i\Vec{k}\cdot\Vec{y}}
    =-\sum_{\Vec{k}\in \mathbb{Z}^3\backslash \{0\}} e^{2\pi i\Vec{k}\cdot\Vec{y}} \,, 
\ee
where $\Vec{y}_{\pi_3}=(y^5,y^6,y^7)$ and we use the discrete Fourier transforms of the delta functions to respect the toroidal periodicity.
From this we deduce that 
\be
\phi_{000} = \sum_{\Vec{k}\in \mathbb{Z}^3\backslash \{0\}} \frac{1}{4\pi^2 k^2}e^{2\pi i\Vec{k}\cdot\Vec{y}} \,, 
\ee
and similarly for the other $\phi_{mnk}$. 
So the Poisson equation in \eqref{P2@} is solved for
\begin{align}
\nonumber 
    \beta(y)&=  \sum_{m_a\in \{0,1\}}\sum_{\Vec{k}\in \mathbb{Z}^3\backslash \{0\}} \frac{1}{4\pi^2 k^2}e^{2\pi i\Vec{k}\cdot(\Vec{y}-\frac{\Vec{m}}{2})} + \text{const}. 
    \\\label{beta-sol}
    &= \sum_{\Vec{k}\in \mathbb{Z}^3\backslash \{0\}} \frac{1}{2 \pi^2 k^2}e^{4\pi i\Vec{k}\cdot\Vec{y}} + \text{const}. 
\end{align}
The notation is $\Vec{m}=(m_5,m_6,m_7)$, $\Vec{k}=(k_5,k_6,k_7)$ 
and $k^2=k_5^2/r^{(0)2}_5+k_6^2/r^{(0)2}_6+k_7^2/r^{(0)2}_7$.

Since \eqref{beta-sol} is not convergent one may wish to regularize it by following \cite{Shandera:2003gx,Junghans,Andriot:2019hay,Andriot:2021gwv}, 
or by simply introducing a hard cut-off on the magnitude of the momenta $\vec k$. 
However, 
to estimate the backreaction of the O-planes we just need the behavior near one of the loci. 
This means we want to evaluate \eqref{beta-sol} at, say, $\Vec y \to0$. 
Clearly, near such point the impact of the other sources can be ignored 
and the divergence will be dominated only by the source at $\Vec y = 0$. 
Therefore, near the source at $\Vec y \to0$, the equation \eqref{P2@} can be approximated by $\nabla^2 \beta \simeq -  \delta(\vec y)$ 
which has the text-book solution $\beta \simeq  r^5 r^6 r^7 / (4 \pi \sqrt{y^2})$. 
This means that near the O-plane we simply have a $1/|y|$ singularity. 
For completeness we can verify this intuitive behavior in the following way. We first define, $\hat{y}_i = y_i/\epsilon $ and $\kappa_{i} = \epsilon k_{i}/r_i$, such that 
\begin{align}
k^2 &= g^{ij} k_i k_j = \frac{1}{\epsilon^2} (\kappa_5^2+\kappa_6^2+\kappa_7^2) \equiv \frac{1}{\epsilon^2} \kappa^2 \,,  \\
r^2 &\equiv g_{ij} \hat{y}^i \hat{y}^j = \epsilon^2 (r_5 y_5^2 + r_6 y_6^2 + r_7 y_7^2) \, . 
\end{align} 
Dropping finite contributions we can write 
\begin{align} 
\beta(\vec{y}) = \sum_{\kappa_i \in (\epsilon/r_i) \mathbb{Z}\backslash \{0\}} \frac{\epsilon^2 }{2 \pi^2 \kappa^2}e^{4 \pi i \Vec{\kappa}\cdot \hat{y}}= \frac{1}{\epsilon} \sum_{\kappa_i \in (\epsilon/r_i) \mathbb{Z}\backslash \{0\}} \frac{r_5 r_6 r_7}{2 \pi^2 \kappa^2}e^{4 \pi i \Vec{\kappa}\cdot \vec{y}} \Delta \kappa_5\Delta \kappa_6\Delta \kappa_7 \, . 
\end{align}
where $\Delta \kappa_i = \epsilon/r_i$. The near-brane limit is captured by sending $\epsilon \to 0$, in which case the sum becomes an integral and we obtain
\begin{align}
\label{correction}
\beta(\vec{y}) \to \frac{r_5 r_6 r_7}{2\pi^2 \epsilon} \int d^3\kappa \frac{e^{4\pi i \vec{\kappa}\cdot \hat{y} }}{\kappa^2} = \frac{1}{4\pi} \frac{r_5 r_6 r_7}{ \epsilon |\hat{y}|} = \frac{1}{4\pi} \frac{r_5 r_6 r_7}{ r} \, . 
\end{align}
Where the first equality follows from recognizing the Fourier transform of the Coulomb potential.\footnote{Note that carrying out this Fourier transform properly also requires the use of a regularization scheme.} This fixes the behavior of $\beta$ near a single O6-plane.

From the $\beta$ behavior derived in \eqref{correction} and the relation in Eq.(\ref{relations}) 
we can see that the first order correction on the fields near the locus of a \textit{single} O6-plane is 
\begin{align} \label{thats}
   \tau &=\tau^{(0)}n^{3/4}-\frac{3}{8\pi r}n^{-1/4} + \mathcal{O}(n^{-5/4})\,, \\
   \label{what}
   w &=w^{(0)}n^{3/4}+\frac{w^{(0)}}{\tau^{(0)}}\frac{1}{8\pi r}n^{-1/4} + \mathcal{O}(n^{-5/4})\,, \\
   \label{she}
   g_{kk}&=g_{kk}^{(0)}n^{1/2}-\frac{r_i^{(0)2}}{\tau^{(0)}}\frac{1}{4\pi r}n^{-1/2}+ \mathcal{O}(n^{-3/2})\,,~~~~~~~~~ k=5,6,7 \\
   \label{said}
   g_{ii}&=g_{ii}^{(0)}n^{1/2}+\frac{r_i^{(0)2}}{\tau^{(0)}}\frac{1}{4\pi r}n^{-1/2}+ \mathcal{O}(n^{-3/2})\,.~~~~~~~~~ i=1,2,3,4  
\end{align}
Near the local sources the $1/|y|$ corrections play against the $n$ suppression, 
but for large enough $n$ they are always subdominant. 
Conversely, 
for any value of $n$ there is always a region close enough to the O-plane where the leading order backreaction dominates.

For the rest of the O6-planes, 
we have that each one of them wraps one $\Psi_i$ four-cycle, 
thus the Bianchi identity can be immediately generalized to 
\begin{align}
    \text{d}F^{(1)}_2&=-2\sum_i^7 \rho_i \,\Phi_i \,,  \label{dF22}
\end{align}
and the source term is 
\begin{align}
    \rho_i = 1 - \frac18 \sum_{m\in\{ 1,2\}}\delta\Big(y^A-\frac{m}{2}\Big)\delta\Big(y^B-\frac{m}{2}\Big)\delta\Big(y^C-\frac{m}{2}\Big)  \,, 
\end{align}
where the combination of $A,B$ and $C$ is given by 
\be
(A,B,C)_i = \left\{(1,2,7),(3,4,7),(5,6,7),(1,3,6),(2,3,5),(1,4,5),(2,4,6)\right\} \,. 
\ee
Then similarly to Eq.(\ref{dF22}) we have 
\begin{align}
    F_2^{(1)}=-2\star_7 \sum_i^7 \Big(\text{d}\beta_i(y^A,y^B,y^C) \wedge \Psi_i\Big)\,, 
\end{align}
where each $\beta_i$ satisfies a condition of the form \eqref{Poisson}. Then the backreaction near each O-plane has an equivalent form as \eqref{correction} and therefore can be controlled for large enough $n$. 
The full form of $\tau^{(1)}, w^{(1)}$ and the metric follow similarly from the equivalent equations to \eqref{relations} 
to find individual contributions of the form \eqref{thats}-\eqref{said} for each three-cycle and adding them together.

\section{Corrections to the effective scalar potential}

\subsection{Corrections in the absence of net D2/O2 charge}

We want to investigate whether the backreaction corrections affect the leading order 3d scalar potential and as a consequence the scale-separation. Considering the metric decomposition in Eq.(\ref{metric}), the dimensional reduction of the ten-dimensional action \eqref{actionTypeII} gives 
\begin{align}
    S_{10}=2\pi\int \text{d}^{3}x\sqrt{g_3}\int \text{d}^{7}y \sqrt{g_7}w^3\Big(\tau^2R_{10}+\mathcal{L}_m\Big)\,.
\end{align}
Here $R_{10}$ is the ten-dimensional Ricci scalar given in Eq.(\ref{Ricci10}) 
and the $\mathcal{L}_m$ the rest of the kinetic and potential terms. 
In order to get the effective 3d action one should integrate over the internal coordinates. 
However, we just need to write down the action from a three-dimensional point of view and study the contribution of the corrections. The 3d effective action is 
\begin{align}
    S_{3}&=\int \text{d}^{3}x \sqrt{g_3}\Big(\tilde{\mathcal{V}}_7R_{3}-V_{3}\Big)\,, ~~~~~ \tilde{\mathcal{V}}_7=\int \text{d}^{7}y \sqrt{g_7}w^3\tau^2\,, \label{reduction}
\end{align}
where the scalar potential takes the form 
\begin{equation}
\begin{aligned}
    V_3&=\int \text{d}^{7}y \sqrt{g_7}w^3\Bigg(-\tau^2R_{7}+6\frac{\tau^2}{w}\nabla_m\nabla^mw +6\frac{\tau^2}{w^2}\nabla_mw\nabla^mw \\
    &-4g^{mn}\partial_m\tau\partial_n\tau+\frac{1}{2}\tau^{2}\vert H_3\vert^2+\frac{1}{2}\vert F_{p}\vert^2-2\mu_6 \sum_i\tau\delta(\pi_i) \Bigg)\,. 
\end{aligned}
\end{equation}
To see the effect of the first order correction we replace the delta function corresponding to the O6-plane by our next-to-leading order solution for the $\text{d}F_2$ Bianchi identity. 
We start from the volume of the wrapped cycle in Eq.(\ref{deltarelation}) which gives 
\begin{align}
    \int \text{d}^7y\sqrt{g_7}\delta(\pi_i)=\int\text{vol}_{\pi_i}\wedge \delta_{i,3}=\frac{1}{2}\int\text{vol}_{\pi_i}\wedge\Big(H_3\wedge F_0 -\text{d}F_2\Big)\,. 
\end{align}
Replacing this into the effective scalar potential gives, after some manipulations, 
\begin{equation}
\label{fotaras2} 
\begin{aligned}
    V_3&=\int \text{d}^{7}y \sqrt{g_7}w^3\Bigg(-\tau^2R_{7}+6\frac{\tau^2}{w}\nabla_m\nabla^mw +6\frac{\tau^2}{w^2}\nabla_mw\nabla^mw-4 g^{mn}\partial_m\tau\partial_n\tau \\
    &+\frac{1}{2}\tau^{2}\vert H_3\vert^2+\frac{1}{2}\vert F_{p}\vert^2 \Bigg)-\mu_6 \sum_i\int\text{vol}_{\pi_i}\wedge\Big(\tau w^3H_3\wedge F_0 +\text{d}(\tau w^3)\wedge F_2\Big)\,. 
\end{aligned}    
\end{equation}
We now want to bring the effective action Eq.(\ref{reduction}) to the Einstein frame. 
To do this we perform the rescaling $g^S_{\mu\nu}=g^E_{\mu\nu}(2\pi\tilde{\mathcal{V}}_7)^{-1/2}$. 
The Einstein-frame scalar potential of the 3d effective theory is 
\begin{align}
    V^E=\frac{V_3}{(2\pi)^2\tilde{\mathcal{V}}_7^{3/2}}\,,
\end{align} 
and we will be ignoring from now on the superscript $E$. 
We can use this form of the effective potential to estimate the impact of the backreaction. 
We will do this by comparing the contributions from the unsmeared terms to the leading order smeared potential.

Let us find the scaling of the smeared potential first. 
At leading order the volume $\tilde{\mathcal{V}}_7$ scales like $\tilde{\mathcal{V}}_7^{(0)}\sim n^{11/2}$ and $\text{vol}_{\pi_i}^{(0)}\sim n$. 
To find the scaling of $R_7$ we need the scaling of the Ricci tensor in \eqref{Ricci1}. 
We see that $R_{mn}\sim g^{(0)rs}\nabla_m\nabla_n g^{(1)}_{rs}+...$ where nabla contains products of the metric and its inverse so it doesn't scale. 
The scaling of the internal Ricci scalar at leading order is $R^{(0)}_7\sim n^{-3/2}$. 
The zeroth order potential (after few integrations by parts) takes the form 
\begin{equation}\label{smearedpotential}
\begin{aligned}
    V_3^{smeared}&=\frac{1}{(2\pi)^2\tilde{\mathcal{V}}_7^{(0)1/2}\tau^{(0)2}}\Bigg(\frac{1}{2}\tau^{(0)2}\vert H_3^{(0)}\vert^2+\frac{1}{2}\sum_{p=0,4}\vert F_{p}^{(0)}\vert^2 \Bigg)n^{-17/4}  \\
    &-\frac{\mu_6}{(2\pi)^2\tilde{\mathcal{V}}_7^{(0)3/2}}\sum_i\int \text{vol}_{\pi_i}^{(0)}\wedge\Big(\tau^{(0)}w^{(0)3}H_3^{(0)}F_0\Big)n^{-17/4}\,. 
\end{aligned}
\end{equation}
We see that the leading order potential scales as $n^{-17/4}$.

We will now estimate the impact of the backreaction by evaluating the scaling of the terms that correspond to the unsmearing corrections 
by inserting the expansions \eqref{expansion0}-\eqref{metricexpansion} in the effective potential. 
First we can check the term that originates from the leading order correction to the last term in \eqref{fotaras2}. 
The leading order in $n$ is 
\begin{align}
\label{V-correction}
    \delta V_3& \ni -\frac{\mu_6}{(2\pi)^2\tilde{\mathcal{V}}_7^{(0)3/2}}\sum_i\int \text{vol}_{\pi_i}^{(0)}\wedge\Big(\text{d}(3\tau^{(0)} w^{(0)2}w^{(1)}+\tau^{(1)}w^{(0)3})\wedge F^{(1)}_2\Big)n^{-21/4}\,.
\end{align}
Note that there are derivatives of the dilaton and the warp factor. 
We see that this correction term is damped faster for large values of $n$ compared to the smeared term, 
thus the potential matches to the smeared one at the leading order, 
{\it assuming} that the formal singularities of the near-brane regions are somehow resolved from string theory. 
Indeed, the formal expression \eqref{V-correction} hides singularities related to the fact that the solution clearly breaks down 
in the regions of the internal space surrounding the O-plane loci because the $1/r$ terms dominate over the $n$ suppression. 
A way to see this is by focusing on the $\pi_3$ four-cycle backreaction in \eqref{V-correction} and estimating the term 
\be \label{V3cor}
\delta V_3^{\pi_3} \sim 
\frac{n^{-21/4}}{\tilde{\mathcal{V}}_7^{(0)3/2}}\int \text{vol}_{\pi_3}^{(0)}\wedge\Big(\text{d}(3\tau^{(0)} w^{(0)2}w^{(1)}+\tau^{(1)}w^{(0)3})\wedge F^{(1)}_2\Big) \,, 
\ee
in the near-brane region. 
At that limit from \eqref{dF2sol}, \eqref{relations}, \eqref{P2@} and \eqref{correction} we have, 
\be \label{nearO6}
\text{near the O6$_\alpha$ central locus:}\ F_2^{(1)} \sim \frac{1}{|\vec{y}|^2}\ 
, \ \ \text{d}F_2^{(1)} \sim \delta(\vec y) \, \Phi_3 \ 
, \ \ 
\frac{\tau^{(1)}}{\tau^{(0)}} \sim  \frac{w^{(1)}}{w^{(0)}} \sim \frac{1}{|\vec y| \, \tau^{(0)}} \,. 
\ee
We can regularize the divergence of the integral in \eqref{V3cor} by excising regions around the O6 locus, which we take to be three-spheres of radius $r_0$ and denote $S_3(r_0)$. Integrating by parts now produces a non-vanishing boundary term. This leads to an estimation of the near-O6-plane backreaction of the form
\be
\begin{aligned}
\label{catastrophe}
\delta V_3^{\pi_3} (\text{O6$_\alpha$ locus}) 
&\sim \frac{n^{-21/4} w^{(0)3} \mathcal{V}^{(0)}_{\pi_3}}{\tilde{\mathcal{V}}_7^{(0)3/2}} \Big( \  \int_{\partial S_3(r_0)} (3\tau^{(0)} \frac{w^{(1)}}{w^{(0)}}+\tau^{(1)}) F^{(1)}_2 \\
& \qquad - \int_{\tilde \pi_3 \backslash S_3(r_0)} (3\tau^{(0)} \frac{w^{(1)}}{w^{(0)}}+\tau^{(1)}) \text{d}F^{(1)}_2 \Big)  \\
&\sim \frac{n^{-21/4} w^{(0)3} \mathcal{V}^{(0)}_{\pi_3}}{\tilde{\mathcal{V}}_7^{(0)3/2}} \frac{6\pi^2}{r_0} \,.
\end{aligned}
\ee
In the last line we used the relations in \eqref{nearO6} and only the boundary contribution survives, 
because $\text{d}F_2^{(1)}$ vanishes outside the excised regions. 
Clearly, the resulting expression depends on $r_0$, and diverges as we try to shrink the excised regions. 
This simply signals the breakdown of the leading order solution near the O6 planes, 
where stringy corrections to the 10d dynamics are expected to appear. 
These corrections, in principle determine a physical value of $r_0$ such that \eqref{catastrophe} would accurately 
capture the contribution to the potential from fields away from the O6 locus. 
Indeed, 
if we require the backreaction to be negligible we need 
\be
\label{r0vsn}
V_3^{smeared} \gg \delta V_3 \ \ \Rightarrow \ \  n^{-17/4} \gg n^{-21/4} r_0^{-1} \ \ \Rightarrow \ \ n \gg r_0^{-1} \,. 
\ee
This suggests that we can have a good approximation of the true solution for distances from the loci much greater than $1/n$.

We can also estimate the backreaction from other terms to see if the $1/n$ estimate for the safety distance from the loci is good enough. 
We can check for example the dilaton term from the first line in \eqref{fotaras2} focusing on the higher order terms  
\be \label{dilcor}
\delta V_3 \ni \frac{1}{(2\pi)^2\tilde{\mathcal{V}}_7^{3/2}}\int \text{d}^{7}y \sqrt{g_7} \, w^3(y) 
\Bigg( -4 \, \partial_m \, \delta \tau \partial^m \delta \tau  \Bigg) \,. 
\ee
Following the same reasoning as before we find for the leading $n$ unsmearing correction 
\be
\delta V_3^{dilaton} (\text{O6$_\alpha$ locus}) \sim n^{-21/4} r_0^{-1} \,, 
\ee
which agrees with \eqref{r0vsn}.

It is however suggested in \cite{Junghans} that for a 10d ``observer'' the backreaction is stronger and would require $r_0 \gg n^{-1/4}$ to be able to safely ignore the unsmearing effect. The argument in \cite{Junghans} for this is to compare for example $\tau^2 |H|^3$ to $(\partial \delta \tau)^2$ and see that one needs $n \gg r^{-4}$. We note that this condition delineates the regions of the internal space where leading order corrections to the 10d solution already give approximately the correct field profiles. 

The purpose of $r_0$, however, is to properly separate out the additional $g_s$ corrections to the 10d solution, over and above the $1/n$ corrections. Thus, the choice of $r_0$ should be determined by the regions where the string coupling becomes large, i.e. $1/n$. It therefore appears that there is a region $1/n < r < 1/n^{1/4}$, where although the 10d equations of motion can be trusted, the resulting $1/n$ expansion of their {\it solution} can not. The contributions to the scalar potential coming from integrating over those regions likely have to be computed to all orders and appropriately resummed.

Meanwhile the degrees of freedom near the O6 locus, i.e. at $r<1/n$, would have to be captured by a strong-coupling description of the O6 planes, as the string coupling truly becomes large in those regions even at leading order in $1/n$. Unfortunately, in the presence of a Romans mass, such a strong coupling description is currently unavailable.

\subsection{Corrections including a net O2/D2 charge contribution}

When there is no net O2/D2 cancellation, such contribution needs to cancel by fluxes in the tadpole/Bianchi. 
Then there is an extra contribution in the potential that comes from the RR field $\vert F_{4}\vert^2=\vert F_{4A}+F_{4B}\vert^2$ 
and has the form 
\begin{align}
\nonumber
    V_3^{extra} &=\frac{1}{(2\pi)^2\tilde{\mathcal{V}}_7^{3/2}}\int_7 w^3\Bigg( F_{4A}\wedge\star_7F_{4B}+\frac{1}{2}F_{4B}\wedge\star_7F_{4B} - 2^{-3}\mu_2\tau \frac{j_7}{\mathcal{V}_7} \Bigg) \, \\
    &=\frac{1}{(2\pi)^2\tilde{\mathcal{V}}_7^{(0)3/2}}\int_7 w^{(0)3}\Bigg( F_{4A}^{(0)}\wedge\star_7F_{4B}^{(0)}n^{-23/4}+\frac{1}{2}F_{4B}^{(0)}\wedge\star_7F_{4B}^{(0)}n^{-25/4} \Bigg)
    \\
    \nonumber 
    & \quad  - \frac{1}{(2\pi)^2\tilde{\mathcal{V}}_7^{(0)3/2}}\int_7 w^{(0)3}\Bigg( 2^{-3}\mu_2\tau^{(0)} \frac{j_7^{(0)}}{\mathcal{V}_7^{(0)}}n^{-7} \Bigg)  \, ,
\end{align} 
since the extra terms scale as 
\be
    F_{4A}\wedge\star_7F_{4B}\sim n^{3/4} \ , \quad 
    F_{4B}\wedge\star_7F_{4B}\sim n^{-1/4} \ , \quad 
    j_7/\mathcal{V}_7\sim n^{-7/4} \, , 
\ee
and indicatively $F_{4A}\wedge\star_7F_{4A}\sim n^{7/4}$. 
The scaling of $F_{4B}$ is dictated by the Bianchi identity \eqref{BI-EXP-EXP} with $H^{(0)}$ scaling as $n^0$. 
Considering the scaling of the extra contributions it seems that neither the O2/D2 contributions 
nor the terms which contain the $F_{4B}$ scale the same way as the potential in Eq.(\ref{smearedpotential}) 
and are subleading at large values of the parameter $n$.

We stress that we do not unsmear the O2-plane here, 
this requires additional analysis which we leave for a future work. 
However, the analysis of \cite{Blaback:2010sj}, where {\it space-filling} localized and smeared O2 sources are compared, 
shows that at least for supersymmetric solutions the backreaction is not expected to lead to inconsistencies.

\section{Outlook}

In this work we have analyzed the backreaction of localized sources on the scale-separated AdS$_3$ N=1 vacua of massive Type IIA supergravity. 
We have found that when one applies the scale separation limit to the various ingredients then 
the corrections from the localized sources can be made arbitrarily small. 
Therefore away from the sources the solution seems to be under control and its uplift to an actual solution of string theory seems plausible. 
Of course, 
unless the O6-plane singularities that we encountered 
can be resolved within string theory the smeared approximation is bound to fail. 
Moreover, our analysis here was only the first step that accounts only for the leading order backreaction, 
and therefore we do not know at this point if some intricate inconsistency can show up at the next order, 
as the AdS conjecture would imply \cite{Lust:2019zwm}. 
One could further check the consistency of the backreacted solutions by matching with 
the supersymmetric analysis of Type II AdS$_3$ vacua performed in \cite{Passias:2020ubv}. 
These questions and checks are left for future work.

One equally interesting question that could be now addressed is the stability of non-SUSY AdS$_3$ flux vacua, 
which should be unstable according to the swampland conjectures \cite{Ooguri:2016pdq}. 
In particular from the supersymmetric construction in \cite{AdS3} one can also find the non-supersymmetric ``skew-whiffed'' AdS$_3$ vacua, 
where the $F_4$ flux has flipped sign. 
For the moment our leading order analysis has not indicated some pathology of such vacua but it may be that by going to 
next to leading order in the backreaction some pathology  may show up thus verifying \cite{Ooguri:2016pdq}. 
For example, four-dimensional ``skew-whiffed'' vacua were studied recently in \cite{Giri:2021eob,Marchesano:2021ycx} and possible instabilities were detected. 
We also leave the analysis of the non-supersymmetric vacua for a future work.

Finally, 
on a more general note, 
the understanding of three-dimensional non-supersymmetric vacua of string theory is interesting on its own right due to the applications in holography, 
but also as a way to scrutinize the 3d swampland. 
A clear classification of classical de Sitter vacua (as is done in 4d \cite{Andriot:2020vlg,Andriot:2022way}) would have its own merits 
and in addition would verify or challenge the conspiracy of string theory against 
de Sitter \cite{Danielsson:2018ztv,Obied:2018sgi,Andriot:2018wzk,Garg:2018reu,Ooguri:2018wrx}. 
First steps in this direction were done in \cite{dS3,Emelin:2021gzx} where smeared sources in Type IIA/B were also used, 
and it would be interesting to see how the unsmearing procedure we discussed here would change these results.

\section*{Acknowledgements} 
We thank Alex Kehagias and Thomas Van Riet discussions. The work of ME and FF is supported by the STARS grant SUGRA-MAX. GT thanks the Department of Physics and Astronomy ``Galileo Galilei'' of the University of Padova for the hospitality.

\appendix

\section{Einstein equations} 
\label{Ap1a}
In this appendix we list some useful formulas and equations of motion of the Type II action in Eq.(\ref{actionTypeII}) with the presence of O6-planes in Eq.(\ref{actionOpDp}). 
For simplicity, and direct comparison to \cite{Junghans}, 
we absorb $N_{\text{O6}}$ in the $\sum_i \delta(\pi_i)$ part and then performing a dimensional reduction down to $d$-dimensions.

For the metric in Eq.(\ref{metric}) we find the Ricci scalar in terms of the warp factor 
\begin{equation}\label{Ricci10}
\begin{aligned}
    {\cal R}_{10}=w^{-2}R_{d}+R_{(10-d)}&-2dw^{-1}\nabla_m\nabla^mw -d(d-1)w^{-2}\nabla_mw\nabla^mw \, .
\end{aligned}
\end{equation} 
Here we define ${\cal R}_{MN}$ to be the Ricci tensor for the 10d string frame metric $G_{MN}$, 
and we use the same notation for the 7d and the 3d counterparts, i.e. ${\cal R}_{\mu\nu}$ and ${\cal R}_{mn}$, 
whereas when we work with the unwarped external or internal space metrics ($g_{\mu\nu}$ and $g_{mn}$ respectively) we use the notation 
$R_{\mu\nu}$ and $R_{mn}$. 
The {\it dilaton} equations of motion for external $d$ and internal $(10-d)$ metric are
\begin{equation}\label{dilatonD}
\begin{aligned}
    0=& -8\nabla^2\tau + 2\frac{\tau}{w^2}R_d
    -\frac{8d}{w}(\partial_mw)(\partial^m\tau)
    -2d(d-1)\frac{\tau}{w^2}\nabla_mw\nabla^mw
    -4d\frac{\tau}{w}\nabla_m\nabla^mw
    \\
    &
    +2\tau R_{(10-d)} 
    -\tau\vert H_3\vert^2 +2\sum_i \delta(\pi_i) \,.
    \end{aligned}
\end{equation}
The variation of the action with respect to the ten dimensional metric $G_{MN}$ in the {\it string} frame gives the following equations of motion
\begin{equation}\label{variationGMN}
\begin{aligned}
   & \tau^2\Big({\cal R}_{MN} - \frac{1}{2}G_{MN}{\cal R}_{10}\Big) 
   +2\tau G_{MN}\Big(\frac{d}{w}(\partial^\mu w) (\partial_\mu\tau) + \nabla^{2}\tau\Big) 
   \\ 
   & + 2(\partial_M\tau)(\partial_N\tau) - 2\tau\nabla_M\nabla_N\tau \\
   &-\frac{1}{2}\tau^2\Big(\vert H_3\vert_{MN}^2 -\frac{1}{2}G_{MN}\vert H_3\vert^2 \Big) 
   -\frac{1}{2}\sum_{p=0}^6\Big(\vert F_p\vert_{MN}^2 - \frac{1}{2}G_{MN}\vert F_p\vert^2 \Big)-\frac{1}{2} T^{loc}_{MN}=0 \, ,
\end{aligned}
\end{equation}
where we have used that 
\begin{align}
    \nabla_M\nabla^M\tau=\frac{d}{w}\partial_{\mu}w\partial^{\mu}\tau + \nabla_m\nabla^m\tau~, 
    \quad \nabla_m\nabla^m\tau=\nabla^2\tau \,, 
\end{align}
and 
\begin{align}\label{tensoras}
    T_{MN}^{loc}=2\tau G_{MN}|_{Op} \delta(\Sigma_{p+1})  = 2\tau  \Pi_{i,MN} \delta(\Sigma_{p+1}) \,. 
\end{align}
We contract Eq.(\ref{variationGMN}) with the 10d metric to find ${\cal R}_{10}$ and plugging this back in Eq.(\ref{variationGMN}) gives 
\begin{equation}\label{tracedD}
\begin{aligned}
   &-\tau^2{\cal R}_{MN} +\frac{d}{4}\frac{\tau}{w}G_{MN}(\partial w)(\partial\tau) 
   +\frac{1}{4}G_{MN}(\tau\nabla^{2}\tau)
   +\frac{1}{4}G_{MN}(\partial_L\tau)(\partial^L\tau)
   +2\tau\nabla_M\nabla_N\tau \\
   & 
   -2(\partial_M\tau)(\partial_N\tau)  +\frac{1}{2}\tau^2\Big(\vert H_3\vert_{MN}^2 -\frac{1}{4}G_{MN}\vert H_3\vert^2\Big) +\frac{1}{2}\sum_{p=0}^6\Big(\vert F_p\vert_{MN}^2 -\frac{p-1}{8}G_{MN}\vert F_p\vert^2 \Big) \\
   &+\frac{1}{2}\Big(T^{loc}_{MN}-\frac{1}{8}G_{MN}T^{loc}\Big)=0\,,
\end{aligned}
\end{equation}
where for $T^{loc}$ we mean the contraction of \ref{tensoras} with the metric of the source worldvolume 
\begin{align}
    T^{loc}=G^{MN}T^{loc}_{MN} \, .
\end{align}
Now we contract \eqref{tracedD} with $g^{\mu\nu}$ and we get 
\begin{equation}
\label{Einstein1D}
\begin{aligned}
   &-\tau^2 \left(w^{-2} R_d - \frac{d}{w} \nabla^2 w - d (d-1) w^{-2} \nabla w \nabla w \right)
   \\
   & 
   +\frac{d^2}{4}\frac{\tau}{w}(\partial w)(\partial\tau) 
   +\frac{1}{4} d (\tau\nabla^{2}\tau)
   +\frac{1}{4} d (\partial_L\tau)(\partial^L\tau) 
   \\
   & 
  +\frac{1}{2}\tau^2\Big( -\frac{1}{4} d \vert H_3\vert^2\Big) 
  +\frac{1}{2}\sum_{p=0}^6\Big(-\frac{p-1}{8} d \vert F_p\vert^2 \Big) \\
   &+\frac{1}{2}\Big(g^{\mu\nu} T^{loc}_{\mu\nu} - \frac{d}{8}T^{loc}\Big)=0\,. 
\end{aligned}
\end{equation}

Note that we have the relation 
\begin{align}\label{RicciTensorInternal}
    {\cal R}_{MN}\Big{\vert}_{M=m,N=n}=R_{mn}-\frac{d}{w}\Big(\partial_n\partial_mw - \partial_sw\Gamma^s_{mn} \Big)=R_{mn}-\frac{d}{w}\nabla_m\partial_nw \,.
\end{align}


\end{document}